\documentclass[onecolumn]{emulateapj}

\shorttitle{On the progenitors of AM CVn stars as LISA sources}
\shortauthors{Liu, Jiang, \& Chen}

\begin{document}


\title{On the progenitors of AM CVn stars as LISA sources: the evolved donor star channel}


\author{Wei-Min Liu$^{1}$, Long Jiang$^{1}$, and Wen-Cong Chen$^{2,1}$}
\affil{$^1$ School of Physics and Electrical Information, Shangqiu Normal University, Shangqiu 476000, China;liuwmph@163.com\\
$^2$ School of Science, Qingdao University of Technology, Qingdao 266525, China;chenwc@pku.edu.cn\\
}



\begin{abstract}
The space gravitational wave (GW) detector Laser Interferometer Space Antenna (LISA) that was planed to launch in the early 2030s is to detect the low-frequency GW signals in the Galaxy. AM CVn stars were generally thought to be important low-frequency GW sources. Employing the MESA code, in this work we calculate the evolution of a great number of binary systems consisting of a white dwarf (WD) and a main sequence (MS) star, and diagnose whether their descendants-AM CVn stars can be visible by the LISA. The simulated results show that the progenitors of these LISA sources within a distance of 1 kpc are WD-MS binaries with a donor star of $1.0-1.4~M_\odot$ (for initial WD mass of $0.5~M_\odot$) or $1.0-2.0~M_\odot$ (for initial WD mass of $0.7~M_\odot$), and an initial orbital period slightly smaller than the bifurcation period. Our simulations also indicate ten verification AM CVn sources can be reproduced by the standard magnetic braking model, and are potential LISA sources. Based on the birthrate of AM CVns simulated by the population synthesis, the birthrate of AM CVn-LISA sources evolving from the evolved donor star channel within a distance of 1 kpc can be estimated to be $(0.6-1.4)\times10^{-6}~\rm yr^{-1}$, and the predicted number of AM CVn-LISA sources is about $340-810$. Therefore, the evolved donor star channel play an important role in forming AM CVn-LISA sources in the Galaxy.
\end{abstract}


\section{Introduction}
Gravitational-wave (GW) era is coming after the double compact object merger events GW150914 and GW170817 were detected by the advanced LIGO and Virgo detectors \citep{abbo16,abbo17}. Laser Interferometer Space Antenna (LISA) is a space-based GW detector, which are planned to launch in the early 2030s \citep{amar17}. Unlike the advanced LIGO and Virgo, the scientific aims of the LISA are to detect low-frequency GW signals among the bands from 0.1 mHz to 0.1 Hz \citep{sluy11,amar17}. The potential LISA sources include ultra-compact binaries in our Galaxy, supermassive black hole mergers, and extreme mass ratio inspirals (EMRIs) \citep{amar07}. For the ultra-compact binaries, the possible sources can be classified into two sorts: the first one is compact detached binaries including double white dwarfs (WDs), WD - neutron star (NS) binaries, double neutron stars, and binary black holes \citep{nele01b,liu10,kor17,bre20,krem17a,krem18,krem19,liu20,taur18,yu15,taur17,lam18,lam19,ses20,lau20,and19,li20}; the second one can be interacting binaries including cataclysmic variables (CVs), AM CVn stars, ultra-compact X-ray binaries (UCXBs) \citep{chen20b}, and compact intermediate-mass black hole X-ray binaries \citep{chen20}.

AM CVn stars (hereafter AM CVns) are ultra-compact binaries where a WD is accreting materials from a semi-degenerate helium-rich star or a WD \citep{warn95,nele05,solh10}. In observations, it is found that the orbital periods of AM CVn stars lie among 5 to 65 minutes and the mass transfer is being driven by GW radiations. Therefore, AM CVns are proposed to be ideal candidates detecting low-frequency GW signals by the LISA in the near future \citep{nele04,roe07c}.

At present, there exist three evolutionary channels toward AM CVns. The first one is the double WDs model where two detached WDs evolve toward short period system via the angular-momentum loss by the GW radiation until the less massive WD starts the roche lobe overflow (RLOF) \citep{nath81,nele01a}. Recently, the systematic works on this channel were explored by the theoretical modeling \citep{gok07,krem15,kal16,krem17a,krem17b}, and the observations \citep{kup20,riv20}. However, \cite{shen15} proposed that these interacting double WDs would probably merge due to dynamical friction within the expanding nova shell, making this formation channel for AM CVns seem impossible. The second one is the helium donor star channel, in which a non-degenerate helium star is transferring matters to a WD, and the system will subsequently reach a minimum period about ten minutes \citep{savo86,iben87}; The third one is the evolved donor star channel where a main sequence donor star has lost most of its hydrogen envelope in the early mass-transfer stage (like a CV stage), and subsequently transfers the He-rich matter onto the massive WD once the orbital period is less than one hour, and the binary system appears as a AM CVns \citep{pods03a,bree12,cart13b}. Recently, Gaia14aae is identified as the first AM CVn star in which the central white dwarf is fully eclipsed by the donor star \citep{gre18b}, and is proposed as an example of the evolved donor star channel for AM CVns.

Based on the double WDs channel and the helium star channel, \cite{nele01a} explored a thorough binary population synthesis (BPS) work on AM CVns according to whether the tidal coupling between the accretor and the orbital motion are efficient (model II), or non-efficient (model I). Their results show that while the birthrate derived by model II are comparable for WD and helium star scenarios, model I tend to select the helium star channel as the dominated contribution for AM CVns. Considering all the cases, they proposed that the birth rate of AM CVns in the Galaxy lie between $(1.1-6.8)\times 10^{-3}~{\rm yr}^{-1}$ and give a space density $\sigma=(0.4-1.7)\times 10^{-4}~\rm pc^{-3}$. However, based on the large data of Sloan Digital Sky Survey (SDSS), the later works by \cite{roe07c} and \cite{cart13a} proposed that the space density of AM CVns are supposed to be much lower than expected by a factor of 2-3 (see also our discussion in Section 4). Recently, using Gaia DR2, \cite{ram18} present the lower limit for the space density of AM CVns to be $\gtrsim 3\times 10^{-7}~\rm pc^{-3}$.

Adopting the evolved donor star channel, \cite{pods03a} carried out a detailed BPS calculations on cataclysmic variables (CVs) that can evolve toward AM CVns, and obtained a birthrate of ultra-compact WD binaries (including AM CVns) to be $(0.5-1.3)\times10^{-3}~{\rm yr}^{-1}$.

Recently, \cite{taur18} investigated a systematic work on the evolution of neutron star-WD binaries as LISA source, and found that the formation and evolution of AM CVns are very similar to those of UCXBs. Obviously, the birthrate and number of AM CVns appearing as LISA sources are greater than those of UCXB-LISA sources in the Galaxy due to the initial mass function. In this paper, we attempt to explore a detailed evolution of AM CVns based on the evolved donor star channel, and evaluate the detectability of AM CVns by the LISA. This paper are organized as follows. The input physics and the stellar evolution code are described in Section 2. In Section 3, we summarize the detailed simulated results. The discussion and summary are presented in Sections 4, and 5.

\section{Evolutionary code}

We carried out binary evolutionary calculations toward AM CVns by the Modules for Experiments in Stellar Astrophysics (MESA, version r8845) \citep{paxt11,paxt13,paxt15}, with MESA SDK for Linux (version 20160129) by \cite{tow16}. The primordial binary is composed by a WD (as a point mass of $M_{\rm WD}$) and a main sequence (MS) companion star (with a mass of $M_{\rm d}$). The chemical composition of the donor star is $X= 0.7, Y= 0.28, Z = 0.02$. Radiative opacities of the donor star are primarily from OPAL \citep{igl93,igl96}, with low-temperature data from \cite{fer05}. The Roche lobe radius of the donor star are computed by using the expression given by \cite{egg83}. Mass transfer rates by the Roche lobe overflowing are determined following the prescription of \cite{rit88}. The orbital angular momentum loss (AML) plays a key role in influencing the formation and evolution of AM CVns. We consider three kinds of AMLs including gravitational radiation \citep{land75}, magnetic braking \citep{rapp83}, and mass loss. During the evolution, once the donor star develops a convective envelope and possesses a radiative core, magnetic braking (with magnetic braking index $\gamma=4$, see also our discussions in Section 4 ) will turn on. Note that many setting in the simulations are similar with \cite{taur18}. In our calculations, two initial WD mass $M_{\rm {WD, i}}$ =0.5, 0.7 $M_\odot$ are used. In the MESA code, $\alpha$ = 0 and $\delta$ = 0 are set which means that during RLOF any wind-mass loss from the donor star or formation of a circumbinary disk is neglected. For $\beta$ factor which represent the mass loss fraction from the WD vicinity are set to be 0.8 and 0.7 for $M_{\rm {WD, i}}$ =0.5, 0.7 $M_\odot$, respectively. The Eddington accretion limit are set to be 2.6$\times10^{-7}\,M_\odot\,{\rm yr^{-1}}$ and 3.6$\times10^{-7}\,M_\odot\,{\rm yr^{-1}}$ for $M_{\rm {WD, i}}$ =0.5, 0.7 $M_\odot$, respectively. The excess matter are assumed to be ejected from the vicinity of the accreting WD, and carry away the specific orbital angular momentum of the WD. In our simulation, the irradiation effect of X-ray luminosities of accreting WD is not considered.  The run will stop until the stellar age is beyond the Hubble time.

Our model assumptions are nearly consistent with \cite{pods03a} including the initial donor star and WD masses (the case that $M_{\rm {WD,i}}=1.0\,M_\odot$ can be ignored due to a low weight), the initial chemical composition, the updated opacities, the mixing-length parameter, and the magnetic braking index ($\gamma=4$). The unique difference between these two works is the $\beta$ factor (the mass loss fraction from the WD vicinity), which was taken to be 1 in \cite{pods03a}. However, the numerical calculations indicate that the change of $\beta$ can hardly influence the orbital evolution of the WD-MS binaries (see also Figure 5, which illustrates the evolutionary tracks in the $P_{\rm orb}-M_{\rm d}$ diagram when $\beta=1$, and 0.7). Because our calculations are based on the evolved donor star model that is same to \cite{pods03a}, it is relatively dependable to employ the simulated birthrate of AM CVns given by \cite{pods03a} in the Section 4.

\begin{figure}
\centering
\includegraphics[scale=0.35]{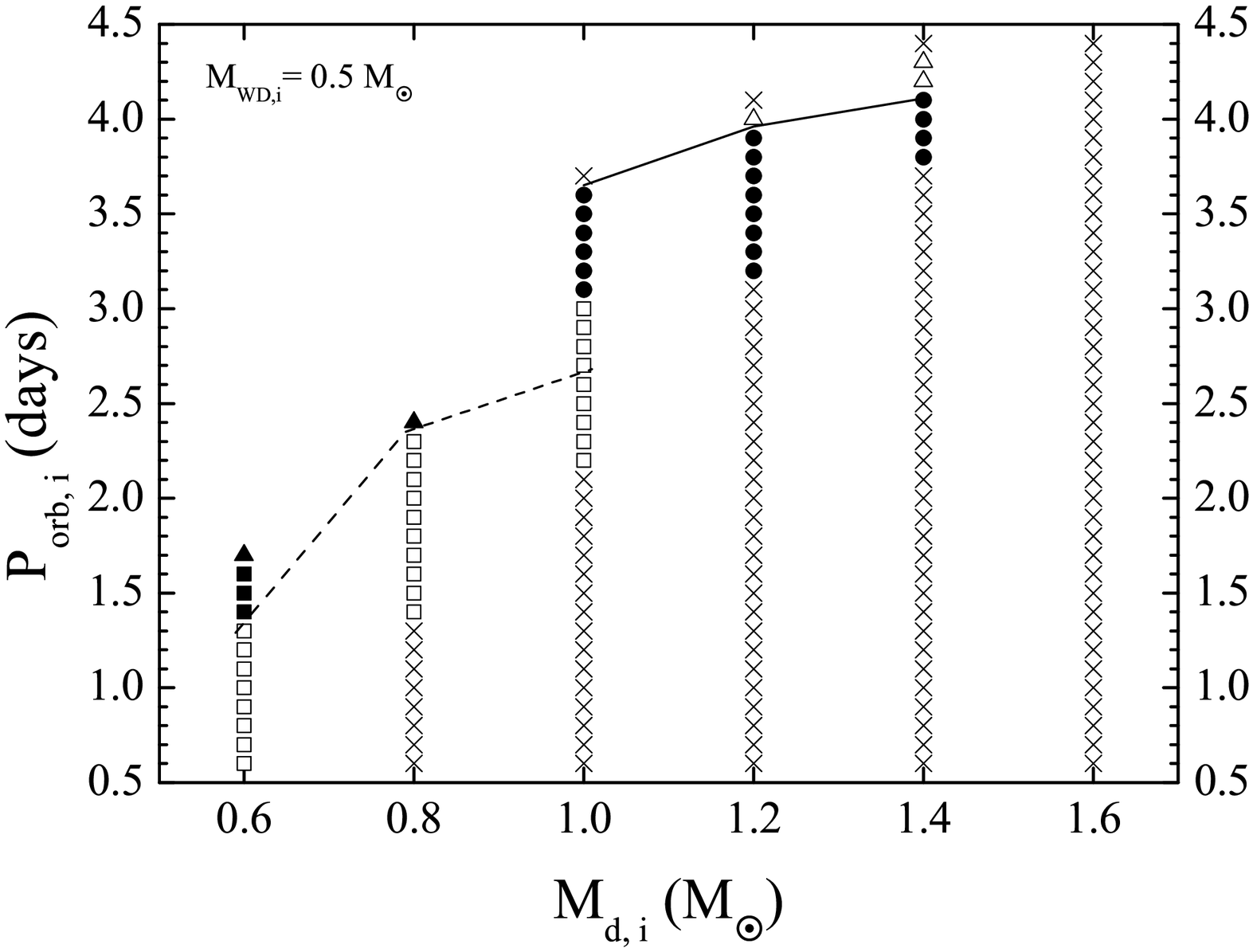}
\includegraphics[scale=0.35]{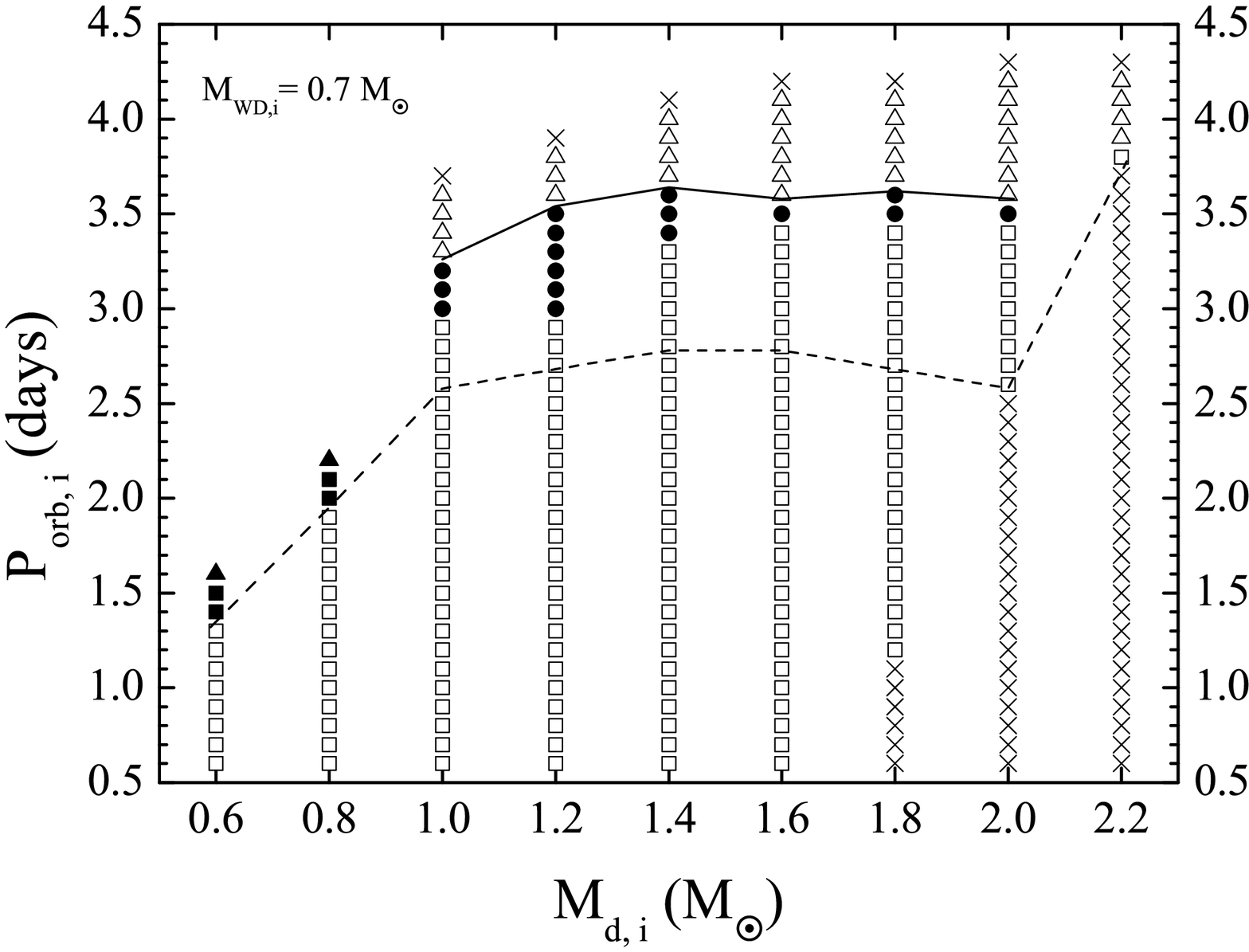}
\caption{Distribution of WD-MS binaries with different evolutionary fates in the initial orbital period vs. initial donor-star mass diagram. The initial WD masses are $M_{\rm {WD,i}}$ = 0.5, and 0.7 $M_\odot$ in the top and bottom panels, respectively. The solid circles represent the WD-MS binaries that can evolve toward AM CVns and can be detected by the LISA (hereafter defined as AM CVn-LISA sources) within a distance $d=1~\rm kpc$. The crosses denote the systems experiencing unstable mass transfer. The open triangles indicate the binaries that evolve toward systems with long orbital periods. The open squares denote the binaries that can evolve toward AM CVns that are invisible by the LISA. Note that the dashed curve denote the boundary line, under which AM CVns cannot be formed because the core hydrogen mass fraction are still larger than 0.4 (we define them as AM CVn-likes).}
\end{figure}

Our calculations show that the descendants of the WD-MS binaries can be detected by the LISA when their orbital period $P_{\rm orb}\lesssim$ 1.5 hours, which is consistent with a GW frequency of 0.4 mHz. Adopting a mission lifetime $T$ = 4 years, the characteristic strain of AM CVns can be calculated by \citep{chen20}
\begin{equation}
h_{\rm c}\approx 3.75\times 10^{-19}\left(\frac{f_{\rm gw}}{1~\rm mHz}\right)^{7/6}\left(\frac{\mathcal{M}}{1~M_{\odot}}\right)^{5/3}\left(\frac{1~\rm kpc}{d}\right),
\end{equation}
where $f_{\rm gw}=2/P_{\rm orb}$ is the GW frequency, $d$ is the distance of the AM CVns. The chirp mass can be expressed as \citep{taur18,bre18}
\begin{equation}
\mathcal{M}=\frac{(M_{\rm WD}M_{\rm d})^{3/5}}{(M_{\rm WD}+M_{\rm d})^{1/5}},
\end{equation}
During the evolution, the corresponding AM CVns are thought as LISA sources once the calculated characteristic strain is larger than the LISA sensitivity,.

\begin{figure}
\centering
\includegraphics[scale=0.35]{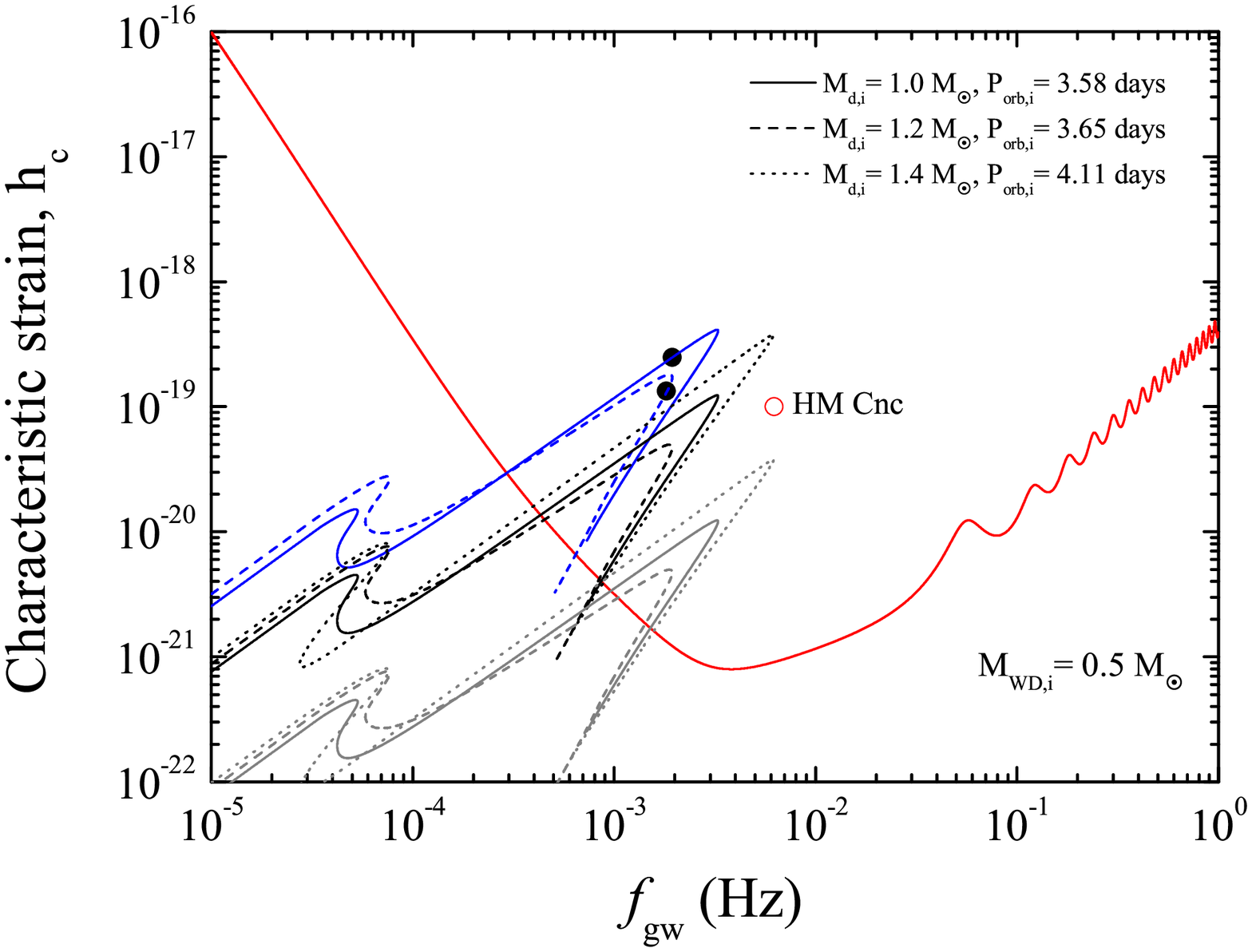}
\includegraphics[scale=0.35]{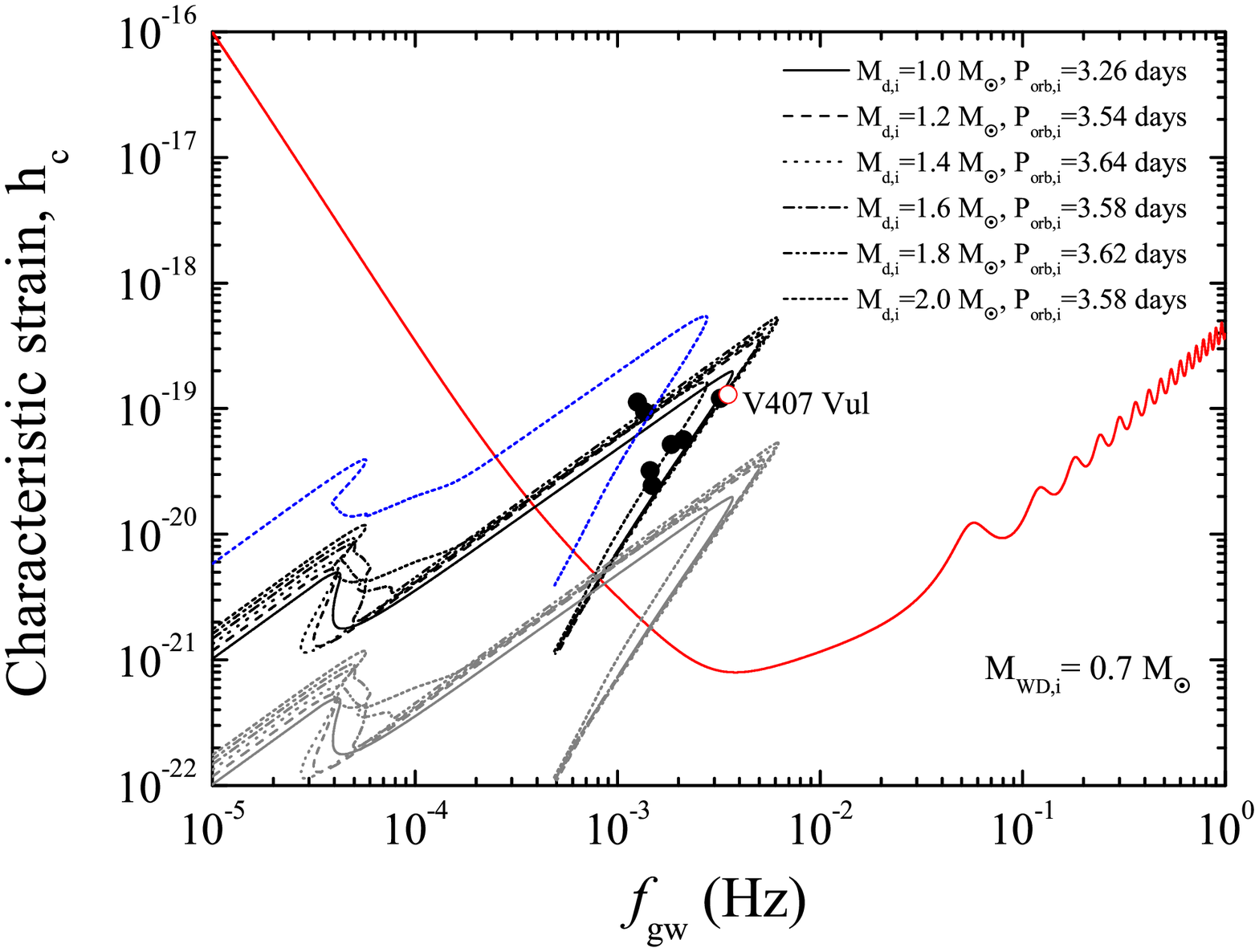}
\caption{Evolution of pre-AM CVns and AM CVns systems in the characteristic strain vs. GW frequency diagram. The red curve denotes the LISA sensitivity curve originated from the numerical calculation. The circles represent the 11 known AM CVn-type verification sources for LISA (HM Cnc, AM CVn and HP Lib in the top panel). The open circles denote the sources HM Cnc, and V407 Vul in the top and bottom panels, respectively. The blue , black, and grey curves correspond to the evolutionary tracks when the distance $d$ = 0.3, 1, and 10 kpc, respectively. }
\end{figure}

\begin{table}
\begin{center}
\caption{Physical Properties of 11 known verification AM CVn stars}
\centering
\begin{tabular}{@{}cccccccccccc@{}}
\hline\hline\noalign{\smallskip}
$\rm Source$ & $d$ &$P_{\rm orb}$ & $M_{\rm WD}$ & $M_{\rm d}$ & $\rm Refs$ \\
        & ($\rm pc$) &($\rm second$) &($M_\odot$) &($M_\odot$) &    \\
\hline\noalign{\smallskip}
HM Cnc &5000 &321.53   &0.55   &0.27   &  1,2\\
V407 Vul &1786  &569.39   &0.8   &0.177   & 3  \\
ES Cet &1584   &620.21  &0.8   &0.161   & 4  \\
SDSS J1351-0643 &1317   &943.84  &0.8   &0.100   & 5 \\
AM CVn &299   &1028.73   &0.68   &0.125   & 6,7  \\
SDSS J1908+3940 &1044   &1085.7   &0.8   &0.085   &8,9  \\
HP Lib &276   &1102.70   &0.64   &0.068   & 10,11  \\
PTF1 J1919+4815 &1338   &1347.35   &0.8   &0.066   & 12 \\
CXOGBS J1751-2940 &971   &1375.0  &0.8   &0.064   & 13 \\
CR Boo&$337^a$   &1471.3  &0.88   &0.066   & 11,14 \\
V803 Cen &$347^a$ &1596.4   &0.98   &0.084   & 11,15\\
\hline\noalign{\smallskip}
\end{tabular}\\
\tablenotetext{}{}{References. [1]\cite{str05}; [2]\cite{roe10}; [3]\cite{ram02}; [4]\cite{esp05}; [5]\cite{gre18a}; [6]\cite{ski99}; [7]\cite{roe06}; [8]\cite{fon11}; [9]\cite{kup15}; [10]\cite{pat02}; [11, $a$]\cite{roe07a}; [12]\cite{lev14}; [13]\cite{wev16}; [14]\cite{pro97}; [15]\cite{roe07b}}
\end{center}
\end{table}

\section{Simulated Results}
Figure 1 shows the evolutionary results of the WD-MS binaries in the initial orbital period versus the donor star mass diagrams. The solid curves represent the bifurcation period \citep{tutu85,pyl88,pyl89}. The bifurcation period is defined as a critical initial orbital period of a binary, above or below which the binary will evolve toward a diverging or converging system. The origin of the bifurcation period comes from the competition between orbital expansion caused by mass transfer from a less massive donor to a more massive accretor, and orbital shrinking due to AML caused by magnetic braking and gravitational radiation \citep[e.g., see Equation 14 in][for a relationship of the parameters]{liu17}. The evolution of a binary with an initial orbital period less than the bifurcation period will be dominated by AML timescale, otherwise the evolution is dominated by the nuclear expansion timescale of the donor star \citep{pyl88}. Therefore, the bifurcation period plays a vital role in determining the evolutionary fates of all types of accreting compact objects \citep[see also][]{ma09,pods02,chen16,jia16,liu17}.  Therefore, above the bifurcation period, the WD-MS binaries would evolve toward wide orbit systems. The dashed curves denote the critical orbital periods, under which the WD-MS binaries would evolve toward AM CVns-like systems including a donor star with a hydrogen mass fraction larger than 0.4. The donor stars in four WD-MS binaries (see also solid triangles) cannot fill the Roche lobes within the Hubble time because of the low donor-star masses and long orbital periods. The solid squares in the two panels represent the WD-MS binaries that cannot evolve into AM CVns or AM CVn-likes in the Hubble time.

The crosses in Figure 1 represent the WD-MS binaries experiencing an unstable mass transfer. The stability of mass transfer is determined by two exponents of donor-star (or Roche lobe) radius to mass as $\zeta_{\rm ad}=(\frac{d\,{\rm ln}R_{\rm d}}{d\,{\rm ln}M_{\rm d}})_{\rm ad}$ and $\zeta_{\rm RL}=\frac{d\,{\rm ln}R_{\rm {RL,d}}}{d\,{\rm ln}M_{\rm d}}$ \citep{sob97,tout97}, where $\zeta_{\rm ad}$ is the adiabatic response of the donor star to mass loss, $R_{\rm d}$, $R_{\rm {RL,d}}$ are the radius of the donor star and its Roche-lobe, respectively. If $\zeta_{\rm RL}>\zeta_{\rm ad}$, the radius of the donor star that is obviously larger than its Roche-lobe radius cannot respond quickly enough to the orbital angular momentum changes, and the donor star deviates from hydrostatic equilibrium. If the initial donor-star masses equal to $1.6~M_{\odot}$ ($M_{\rm WD,i}=0.5~M_{\odot}$) or $2.2~M_{\odot}$ ($M_{\rm WD,i}=0.7~M_{\odot}$), the WD-MS binaries would experience an unstable mass transfer due to a high mass ratio, and are impossible to form AM CVns. Furthermore, the unstable mass transfer also occurs if the initial orbital period is too small (the separation of the binary will continuously shrink) or too large (the donor develops a deep convective envelope prior to the mass transfer). As a result, the mass transfer proceeds on a dynamical timescale and becomes unstable \citep{taur06,paxt15}. This runaway mass transfer event (the mass transfer rates are approximately in the range of $10^{-4}-10^{-3}M_\sun~{\rm yr}^{-1}$) causes the timestep of the calculation to quickly decreases below the limit setting in the MESA, and the calculation ceases. Subsequently, the systems are thought to enter a common envelope stage. Therefore, the initial donor-star masses of WD-MS binaries that can evolve toward AM CVn-LISA source within a distance of 1 kpc are in the range of $1.0-1.4~M_{\odot}$ ($M_{\rm WD,i}=0.5~M_{\odot}$) or $1.0-2.0~M_{\odot}$ ($M_{\rm WD,i}=0.7~M_{\odot}$), and the initial orbital periods are in a narrow range of 3.1 - 4.1 days ($M_{\rm WD,i}=0.5~M_{\odot}$) or 3.0 - 3.6 days ($M_{\rm WD,i}=0.7~M_{\odot}$), in which the upper boundaries are the bifurcation periods.

\begin{figure}
\centering
\includegraphics[scale=0.35]{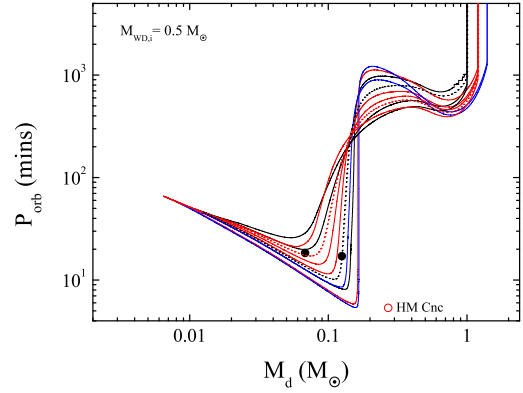}
\includegraphics[scale=0.35]{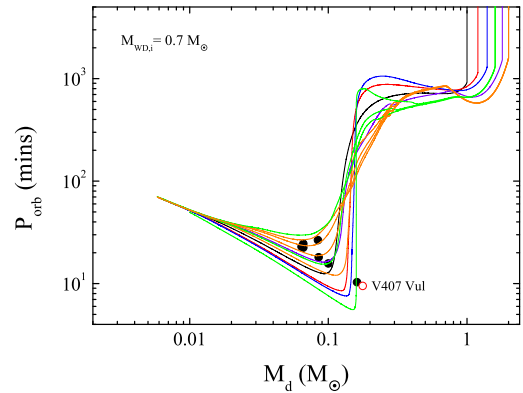}
\caption{Evolutionary tracks of WD-MS binaries in the $P_{\rm orb}-M_{\rm d}$ diagram. The circles denote 11 known AM CVns type verification sources for the LISA. The black, red, blue, green, violin, and orange curves correspond to the binaries with different initial orbital periods and initial donor-star masses $M_{\rm {d,~i}}$ = 1.0, 1.2, 1.4, 1.6, 1.8 and 2.0 $M_\odot$, respectively.}
\end{figure}

\begin{figure}
\centering
\includegraphics[scale=0.35]{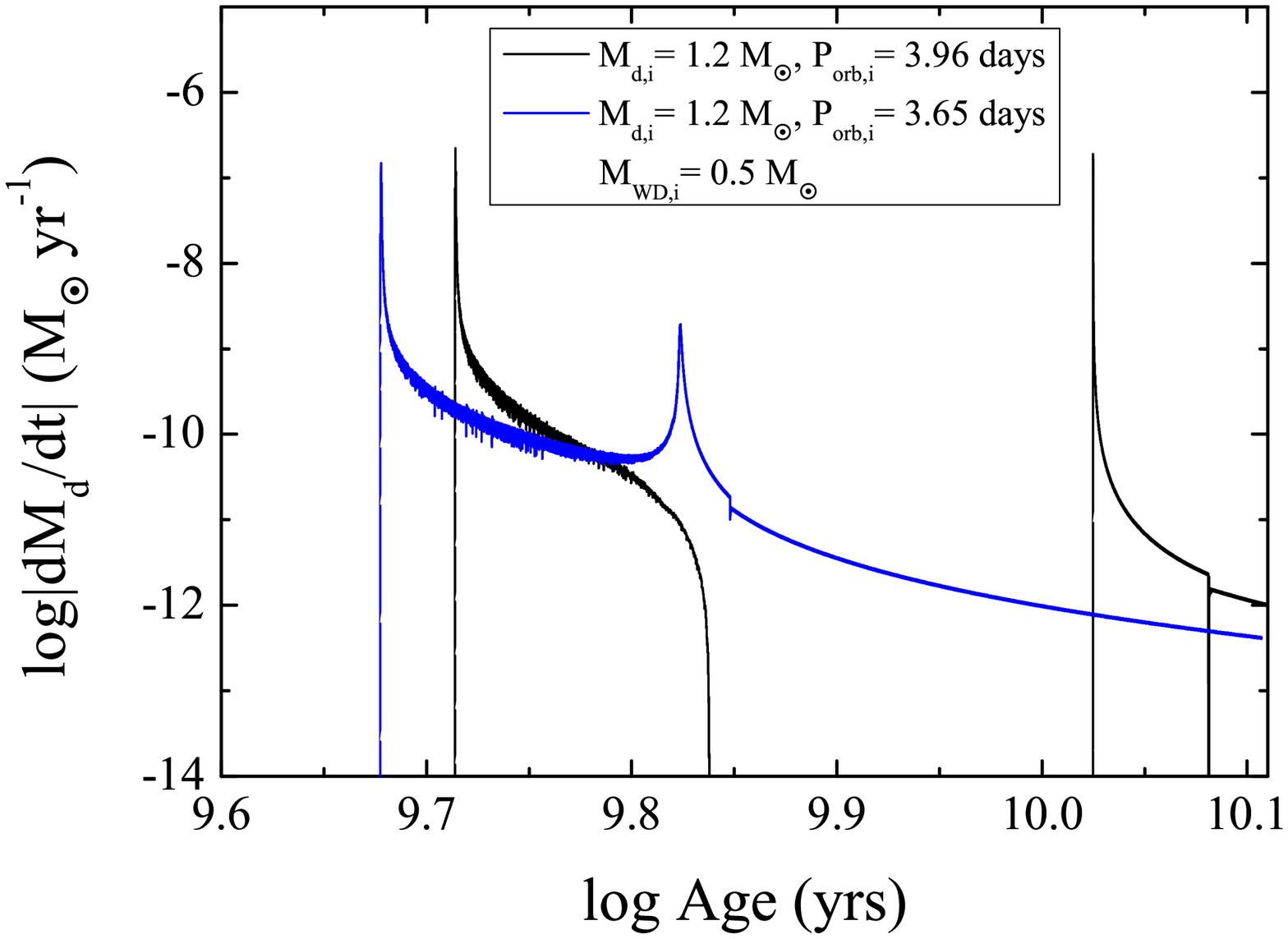}
\includegraphics[scale=0.35]{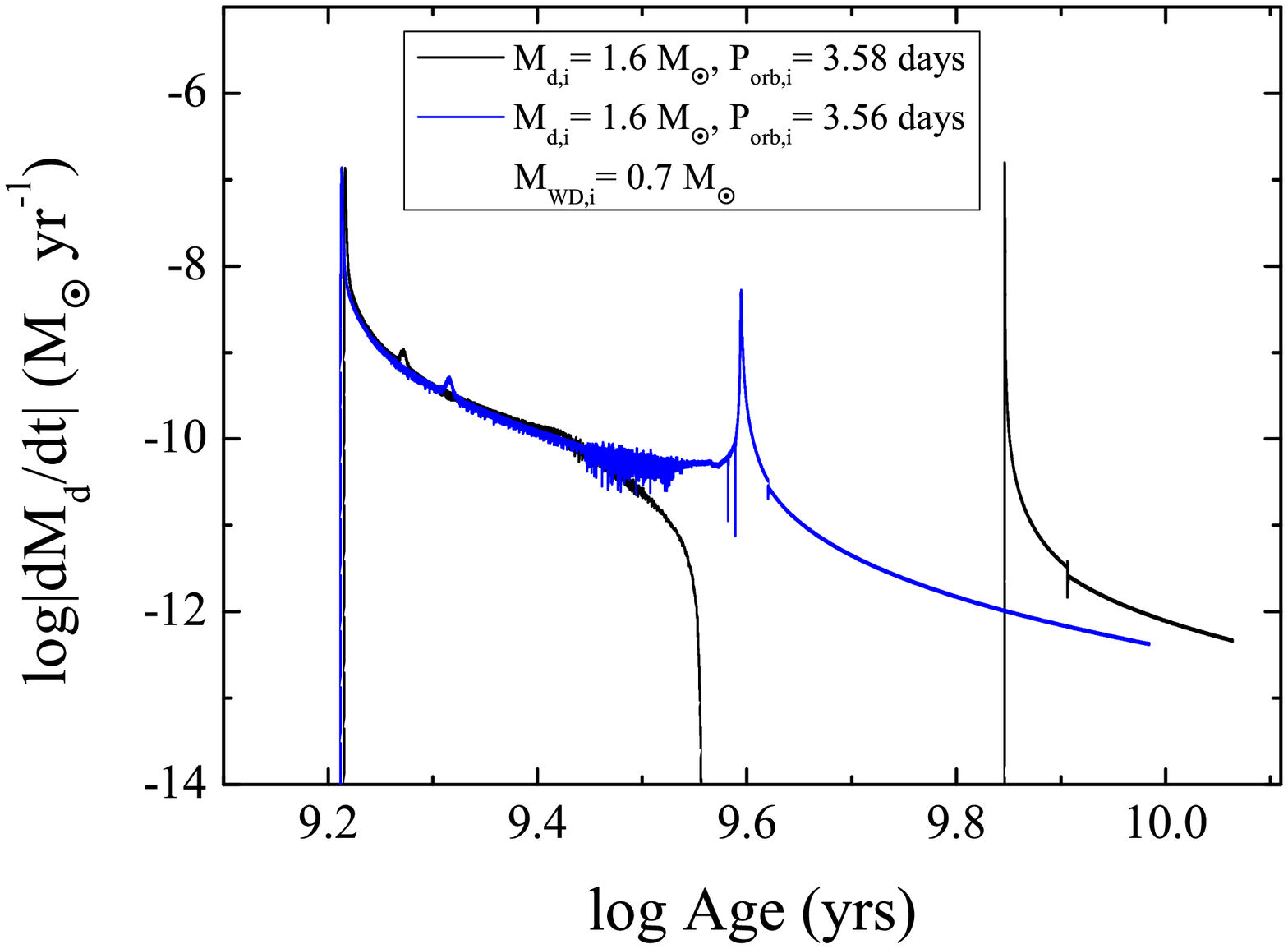}
\caption{Mass-transfer rate of the WD-MS binaries as a function of the stellar age.}
\end{figure}

Figure 2 shows the final evolution of WD-MS binaries with different initial donor-star masses and initial orbital periods in the characteristic strain versus GW frequency diagrams. Considering the possible differences of the initial WD masses, they are assumed to be $M_{\rm {WD,i}}$ = 0.5 and 0.7 $M_\odot$ in the top and bottom panels, respectively. Even if $d=10~\rm kpc$, all our simulated pre-AM CVns and AM CVns can be visible as LISA sources (see also the grey curves in the top and bottom panels), implying AM CVns are important LISA sources in the Galaxy. Table 1 lists the observed and derived parameters of eleven verification AM CVns \citep{kupf18}. We then obtain the characteristic strain of the ten AM CVns according Equation (1). Because the distances of sources AM CVn and HP Lib with low-mass WD are 0.299 and 0.276 kpc, we plot the evolutionary tracks with the distance $d=0.3~\rm kpc$ by the blue curves (the solid and dash lines are for $M_{\rm {donor,i}} = 1.0~M_\odot,~P_{\rm {orb,i}}=3.58$ days and $M_{\rm {donor,i}} = 1.2~M_\odot,~P_{\rm {orb,i}}=3.65$ days respectively) in the top panel, which can fit these two sources very well. It seems hard for our model to reproduce the verification source HM Cnc due to a relative low primary mass ($\approx0.55~ M_\odot$) and a long distance (5 kpc). The remaining sources with relatively high-mass WD can be reproduced by WD-MS binaries with high initial mass WD in a distance of 1 kpc or 0.3 kpc (the blue short dash line for $M_{\rm {donor,i}}~=~2.0~M_\odot$, and $P_{\rm {orb,i}}=3.58$ days).

\begin{table*}
\begin{center}
\caption{Selected Evolutionary Properties for AM CVns and their progenitors for different initial donor star masses and initial orbital periods. \label{tbl-2}}
\begin{tabular}{@{}ccccccccccccc@{}}
\hline\hline\noalign{\smallskip}
$M_{\rm d,i}$ & $M_{\rm WD,i}$& $P_{\rm orb,i}$ &  $t_{\rm RLOF}$ &$t_{\rm deta}$ &$P_{\rm deta}$ &$M_{\rm d,deta}$&$t_{\rm AM CVn}$ &$P_{\rm AM CVn}$&$P_{\rm min}$ & $f_{\rm i,LISA}$ &$\bigtriangleup t_{\rm LISA}$ \\
 ($ M_{\odot}$) &  ($ M_{\odot}$)  &  (days)  & (Gyr)   & (Gyr)  & (days)    &  ($ M_{\odot}$ )  &  (Gyr)  & (mins)& (mins) & (mHz)& (Myr)\\
\hline\noalign{\smallskip}
1.2 & 0.5 & 3.96 & 5.174 & 6.890 & 0.287 & 0.163 & 10.59 &13.98 & 5.86 & 0.40 &559.5\\
1.4 & 0.5 & 4.11 & 2.796 & 4.472 & 0.329 & 0.166 & 11.45 &10.52 & 5.38 &0.39 &572.8\\
\hline\noalign{\smallskip}
1.2 & 0.7 & 3.54 & 5.064 & 7.606 & 0.248 & 0.162 & 9.08 &31.48 & 6.68 & 0.37 &569.5\\
1.4 & 0.7 & 3.64 & 2.711 & 5.382 & 0.326 & 0.167 & 10.90 &12.05 & 5.49 & 0.37 &573.9\\
1.6 & 0.7 & 3.58 & 1.641 & 3.605 & 0.254 & 0.155 & 7.02 &14.40 & 6.34 & 0.37 &589.4\\
1.8 & 0.7 & 3.62 & 1.162 & 2.803 & 0.274 & 0.164 & 8.44 &7.34 & 5.36 &0.36 &579.1\\
\hline\noalign{\smallskip}
\end{tabular}\\
\tablenotetext{}{}{The columns name list (in order): the initial donor-star mass, the initial WD mass, the initial orbital period, the stellar age at the onset of RLOF, the stellar age and the orbital period including the donor star mass when the system become detached, the stellar age and the orbital period when the system appears as an AM CVn, the minimum orbital period, the initial GW frequency when the system start to be detected by the LISA within a distance of 1 kpc, and the
timescale that the binary appears as LISA source.}
\end{center}
\end{table*}

To compare our simulated results with the observations, we plot the evolutionary tracks of WD-MS binaries with different WD and donor-star masses in the orbital period versus donor-star mass plane in Figure 3. The open circles denote the eleven known verification AM CVns sources in Table 1. The curves plotted by the black, red, blue, green, violin and orange colors correspond to donor-star masses $M_{\rm {d,i}}$ = 1.0, 1.2, 1.4, 1.6, 1.8 and 2.0 $M_\odot$. Different curves with same color represent the evolutionary tracks with different initial orbital periods and same donor-star mass. Nine sources among eleven verification AM CVns sources can be well reproduced by the standard magnetic braking description given by \cite{rapp83} apart from HM Cnc and V407 Vul. Note that the two dashed curves represent the same systems in the top panel in Figure 2, which can successfully fit the observations of two sources AM CVn and HP Lib with low-mass WDs.

To show the evolutionary history of WD-MS binaries, Figure 4 plots their evolution in the mass transfer rate versus the stellar age diagram. The top panel illustrates the evolutionary examples of a WD-MS binary with $M_{\rm {WD,i}} = 0.5~ M_\odot$ and $M_{\rm {d,i}}$ = 1.2 $M_\odot$. When the initial orbital period $P_{\rm {orb,i}}$ = 3.96 day, the WD-MS binary will experience three stages including CV, pre-AM CVns (or post-CV), and AM CVn stages (see also Table 2). After 5.174 Gyr of nuclear evolution, the donor star fills its Roche lobe and begins a mass transfer, and the accreting WD appears as a CV. At the stellar age $t=6.89$ Gyr, the CV evolves into a detached pre-AM CVn with $M_{\rm {d}}$ = 0.163 $M_\odot$ (the donor star only remains a low-mass He core) and $P_{\rm {orb}}$ = 0.287 day due to the Roche-lobe decoupling. Subsequently, a low-mass He WD is firstly formed after a $\sim 2~\rm Gyr$ contraction stage, and then starts a cooling phase \citep{istr14a}. With the spiral in due to the AML driven by GW radiation, low-frequency GW signals are emitted from the pre-AM CVns \citep{taur18}. At $t=10.59~\rm Gyr$, the low-mass He WD fills its Roche lobe, and triggers the second mass transfer stage when $P_{\rm {orb}}$ = 13.98 minutes. In this stage, the compact WD binary is observed as an AM CVn. Finally, the donor star evolves into a planet-like donor star with a mass of $7.56\times10^{-3}~M_{\odot}$ \citep{taur18}. Similar to \cite{chen20b}, it is very sensitive to the initial orbital period whether the WD-MS binaries can evolve into a detached pre-AM CVns. When the initial orbital period is 3.65 days, the system can still form AM CVns, while it always experiences mass transfer (see also the blue curve in the top panel of Figure 4) without a Roche-lobe decoupling stage. If the initial orbital periods are obviously less than the bifurcation periods, the WD-MS binaries would directly evolve into AM CVns without experiencing a detached pre-AM CVns stage \citep{chen20b}. For a WD-MS binary with $M_{\rm {WD,i}} = 0.7~ M_\odot$ and $M_{\rm {d,i}}$ = 1.6 $M_\odot$, the evolutionary tracks display a similar tendency (see also the bottom panel of Figure 4).

Table 2 lists some main evolutionary parameters of WD-MS binaries that can evolve into detached pre-AM CVns, which would appear as the LISA sources within a distance $d=1~\rm kpc$. Although the accreting objects of AM CVns and UCXBs are different, the whole evolutionary process are very similar because of the same donor stars. The He WD masses in the pre-AM CVns lie within a narrow range of $0.155-0.167~M_{\odot}$, which is similar to the results of AM CVns and UCXBs given by \cite{taur18} and \cite{chen20b}. It is tightly related to the accretion efficiency, and magnetic braking index $\gamma$ whether a WD-MS binary would evolve toward a detached pre-AM CVn. In UCXBs, there exist a relatively wide range of initial orbital periods for NS-MS binaries that can evolve toward detached pre-UCXBs if a high magnetic braking index $\gamma=5$ were adopted \citep{istr14b,seng17}. It seems that the timescale that AM CVns appear as LISA sources are one order of magnitude higher than those for UCXBs in \cite{chen20b}. However, their timescales that UCXBs appear as LISA sources are based on a detection distance of 15 kpc.
\section{Discussion}
Based on a binary-population synthesis (BPS) simulations with a grid of 120 binary
evolution sequences performed by a detailed stellar evolution models, \cite{pods03a}
obtained Galactic birthrates of AM CVns (including AM CVn-likes) evolving from the evolved donor star channel as $\mathcal{R}_{\rm AM}=(0.5-1.3)\times10^{-3}~{\rm yr}^{-1}$, which depends on the efficiency of magnetic braking and the common envelope
efficiency parameters (see also their Table 2). In our parameter space, all AM CVns (see also solid circles and open squares in Figure 1 ) have probabilities $P\approx40\%$ ($M_{\rm WD,i}=0.5~M_{\odot}$) and $P\approx10\%$ ($M_{\rm WD,i}=0.7~M_{\odot}$) to evolve into AM CVn-LISA sources (see also solid circles in Figure 1) within a distance of 1 kpc.

If the AM CVns satisfy a uniform distribution in the Galactic disk, the birthrate of AM CVn-LISA sources
within a distance between 0 kpc to 1 kpc can be written as
\begin{equation}
\mathcal{R}_{0,1}=\mathcal{R}_{\rm AM}P/225,
\end{equation}
here the radius and the scale height of the Galaxy
are assumed to be 15 and 1 kpc.

Adopting a mean probability $P=25\%$, and $\mathcal{R}_{\rm AM}=(0.5-1.3)\times10^{-3}$ \citep{pods03a}, then $\mathcal{R}_{0,1}\approx(0.6-1.4)\times10^{-6}~\rm yr^{-1}$. This birthrate is approximately one order of magnitude higher than that of UCXB-LISA sources within a distance of 1 kpc \citep{chen20b}. Taking a rough timescale that AM CVns appear as LISA sources $\bigtriangleup t_{\rm LISA}\approx560~\rm Myr$, the number of AM CVn-LISA sources within a distance of 1 kpc $N_{0,1}=\mathcal{R}_{0,1}\bigtriangleup t_{\rm LISA}=340-810$. If the space distribution of AM CVns is the same to UCXBs, the birthrate of AM CVn-LISA sources from 0 kpc to 15 kpc $\mathcal{R}_{0,15}\approx6.5\mathcal{R}_{0,1}\approx(0.4-1)\times10^{-5}~\rm yr^{-1}$ (see also equation 12 in Chen et al. 2020 ). Therefore, the number of AM CVn-LISA sources evolving from the evolved donor star channel is approximately $2200-5200$ in the Galaxy.

Based on population synthesis simulations, \cite{nele01a} proposed a space density $\sigma=(0.4-1.7)\times 10^{-4}~\rm pc^{-3}$ for AM CVns, while this estimation was very uncertain because of dependance on the model parameters. Adopting the large scale SDSS data, \cite{roe07c} obtained a space density $\sigma=(1-3)\times 10^{-6}~\rm pc^{-3}$. Subsequently, \cite{cart13a} derived a most reliable estimation of space density $\sigma=(5\pm3)\times 10^{-7}~\rm pc^{-3}$ by a significantly
expanded SDSS sample. There, we can estimate the number of AM CVns within a distance of 1 kpc to be $N=2\pi R^{2}z\sigma=3140\pm1880$. Assuming that the probability of AM CVns that can be detectable by the LISA is similar to our simulation ($P=25\%$), the total number (including the double WDs, the He star, and the evolved donor star channels) of AM CVn-LISA source within a distance of 1 kpc in the Galaxy is $N_{\rm AM, LISA}=790\pm470$. Therefore, our estimated number ($340-810$) of AM CVn-LISA sources within a distance of 1 kpc occupies a fraction of $\sim1/4-2/3$, implying that the evolved donor star channel donates an important contribution on the formation of AM CVn-LISA sources.

Considering two formation routes of AM CVns including double WD channel and He donor star channel, \cite{nele04} found that several thousand AM CVns are potential LISA sources in the Galaxy. Recently, \cite{krem17a} explored the long-term evolution of interacting double WDs involving both direct-impact and disk accretion. By a galactic population synthesis, they predicted that $\sim$ 2700 mass-transferring double WDs can be detectable by the LISA with signal to noise ratio (S/N) $>$ 5 in the Galaxy. Meanwhile, \cite{brow20} found that the merger rate of the observed He + CO WD binaries exceeds the birth rate of AM CVns by a factor of 25, i.e. the majority of He + CO WD binaries would experience unstable mass transfer and merge, and are difficult to evolve toward AM CVns. Therefore, the maximum number of AM CVn-LISA sources in the Galaxy evolving from the double WDs channel should be several thousand. As a result, the predicted number (2200-5200) by the evolved donor star channel is indeed comparable with the double WDs channel, and provides a considerable contribution on the AM CVn-LISA sources in the Galaxy.

\begin{figure}
\centering
\includegraphics[scale=0.35]{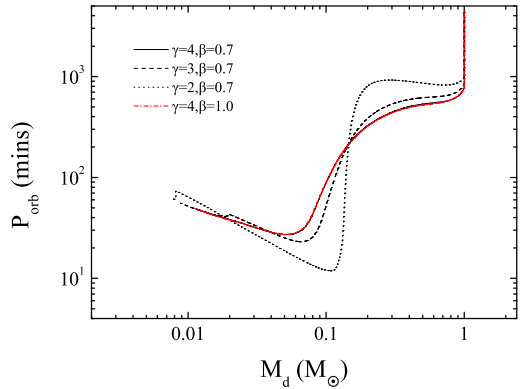}
\caption{Evolutionary tracks of WD-MS binaries with $M_{\rm WD,i}=0.7~M_{\odot}$, $M_{\rm d,i}=1.0~M_{\odot}$, and $P_{\rm orb,i}=3.0~\rm days$ in the $P_{\rm orb}-M_{\rm d}$ diagram.}
\end{figure}

There exist some uncertainties in our estimations for the birthrate and number of AM CVn-LISA sources evolving from the evolved donor star channel. The first uncertainty is AML mechanism by the magnetic braking, which play an important role in determining the evolutionary fates of WD-MS binaries. In the CV or pre-AM CVn stage with orbital periods greater than 3 hours, the AML is dominated by the magnetic braking. Comparing with the magnetic braking, the influence of the mass loss can be ignored (see also Figure 5). However, the gravitational radiation dominates the orbital evolution of the CVs or pre-AM CVns with orbital periods less than 2 hours. Therefore, the standard magnetic braking model is successful to account for the period gap (2-3 hours) of CVs \citep{rapp83}. The numerical simulations by the MESA indicate that a low magnetic braking index (e.g. $\gamma=2,3$) would induce a relatively small minimum orbital period (see also Figure 5), and cause the initial parameter space in Figure 1 slightly move downward. As a result, the birthrate and number of AM CVn-LISA sources would accordingly alter. However, it is successful for the standard magnetic braking with $\gamma=4$ to reproduce the verification AM CVns sources. Therefore, our estimation for the birthrate and number of AM CVn-LISA sources remains fairly reliable.

The second uncertainty is the space density of AM CVns, which would determine the contribution of the evolved donor star channel on the AM CVn-LISA sources. The earliest space density estimation for the AM CVns in the Galaxy from the observations is $\sigma=3\times 10^{-6}~\rm pc^{-3}$ \citep{warn95}. Using the He emission dominated spectra, six new AM CVns were discovered in the SDSS spectroscopic data base \citep{ande05,ande08,roe05}. By calibrating the simulations from BPS, a local space density $\sigma=(1-3)\times 10^{-6}~\rm pc^{-3}$ was obtained \citep{roe07c}. However, the reliability of this result was limited due to a small sample size. Using the latest photometric data base of SDSS, \cite{cart13a} explored 2000 candidates that should include most AM CVns ($\sim50$) in the SDSS to derive an observed space density as $\sigma=(5\pm3)\times 10^{-7}~\rm pc^{-3}$. The accurate distances are very important in modelling the spatial distribution and space density of the AM CVns. Using the distances from Gaia Data Release 2 to the known AM CVns, \cite{ram18} presented a lower limit on the space density to be $\gtrsim 3\times 10^{-7}~\rm pc^{-3}$. Obviously, this space density limit is compatible with that given by \cite{cart13a}. Meanwhile, \cite{ram18} found that the mass transfer rate in most sources among 15 AM CVns are greater than predicted by standard tracks, implying the majority donor stars in AM CVns population are not fully degenerate. This evidence also indicates that the evolved donor star channel cannot be negligible in forming AM CVns population.
\section{Summary}

In this work, we investigate the formation and evolution of AM CVns producing by the WD-MS evolutionary channel, and diagnose the detectability of these AM CVns as LISA sources. Our main conclusions are as follows:

(1) The initial donor-star masses and initial orbital periods of WD-MS binaries that can evolve toward AM CVn-LISA sources within a distance of 1 kpc are in the range of $1.0-1.4~M_\odot$ and $3.1-4.1~\rm days$, and $1.0-2.0~M_\odot$ and $3.0-3.6~\rm days$ when the initial WD masses are 0.5, and $0.7~M_\odot$, respectively. In our investigated parameter space, a fraction 40\% and 10\% of AM CVns can evolve toward AM CVn-LISA sources that can be visible within a distance of 1 kpc for $M_{\rm {WD,i}} = 0.5$, and  $0.7~M_\odot$, respectively.

(2) The progenitors of all AM CVn-LISA sources should have an initial orbital period slightly smaller than the bifurcation period. If the initial orbital periods are much smaller than the bifurcation periods, the relevant AM CVns would be invisible by the LISA. If the initial orbital periods are approximately equal to the bifurcation periods, the WD-MS binaries would experience three stages including CVs, the detached WD-He WD binaries, and AM CVns. In this case, the AM CVns would emit relatively high frequency GW signals, which can be detected by the LISA even if for a long distance of 10 kpc.

(3) In the detached pre-AM CVns, the He WD masses are in a narrow range of $0.155-0.167~M_{\odot}$, which can be used to constrain the primary WD masses \citep{taur18}.

(4) The standard magnetic braking model given by \cite{rapp83} can reproduce the orbital periods and the derived donor-star masses of ten verification AM CVns sources.

(5) Based on a birthrate of AM CVns given by the BPS simulations \citep{pods03a}, the birthrate of AM CVn-LISA sources evolving from the evolved donor star channel within a distance of 1 kpc is estimated to be $(0.6-1.4)\times10^{-6}~\rm yr^{-1}$, and the relevant number of AM CVn-LISA sources is about $340-810$. Comparing with the derived space density by significantly expanded SDSS sample, the evolved donor star channel contributes a considerable fraction of AM CVn-LISA sources.

\acknowledgments {We are grateful to the referee for her/his valuable comments that have led to the improvement of the manuscript. We would also like to thank Professor Xiang-Dong Li for helpful discussions. This work was partly supported by the National Natural Science Foundation of China (under grant Nos.11733009 and U2031116), the Program for Innovative Research Team (in Science and Technology) at the University of Henan Province, and the Research Start-up Funding Project of Shangqiu Normal University.}

\end{document}